# Comparative Survey of Cyber-Threat and Attack Trends and Prediction of Future Cyber-Attack Patterns

Uwazie Emmanuel Chinanu[1*] and Oluyemi Amujo[2]

[1]Centre for Cyberspace Studies, Nasarawa State University, Keffi, Nigeria

[2]University of Abuja, Abuja, Nigeria

E-mail: uwazieemmanuel@yahoo.com
E-mail: oluyemiamujo@gmail.com

**Abstract:** This paper presents a comparative survey of cyber-threat and attack trends starting from 2010 till date. Cyber security breaches are constantly on the rise with huge uncertainty and risks. The trend is causing rife globally because of its consequences to national security and economy. With diverse interests and motivations for various categories of threats and attacks, we carried out a comparative survey and analysis of security breaches to unravel the patterns and predict what will shape future security challenges. The diversity of attacks and growing state actors' involvement without any sort of regulation is making cyber weapons attractive to the states. States are leveraging the anonymity and attribution flaws to hit hard on perceived adversaries thereby complicating the cyber security equation.

## I. INTRODUCTION

With the increasing dependence on the internet by individuals, industries, academic institutions and government agencies, the issue of cyber-attack as become everyone's business. While the cyber space is relied on for entertainment, business, education and administration amongst other purposes, many users –individuals, groups and nation states- now take advantage of this dependence to fulfil their own malicious intents [1]. Different attacks now occur on the internet on a regular basis. While some of them are similar in purpose and mode of operation, others tend to be of a different stalk. Some attacks are re-occurrences of past attacks - with different targets and tools used for operation, while other attacks are entirely new. Knowledge of the sources of these attacks, system vulnerabilities that paved way for the attacks, and the method used in carrying out the attacks is crucial for mitigating the re-occurrences of similar attacks in the future [2]. Comparative Survey of Cyber-Threat and Attack Trends and Prediction of future Cyber-attack Patterns is for every stake holder of the cyber space who is concerned about being secured against cyber-threats. The study high lights real case studies of attacks, identifies their motivations, tools or techniques used and the targets they exploited. These events are statistically analysed to predict future attacks, in a bid to inform internet users to be better prepared for them. The rest of this paper is organised into different sections. Section II reviews various literature that explain the diverse cyber-threats and attacks that exist and section III identifies the various cyber-threats and attacks that have plagued systems since the year 2012 till date. Section IV describes a methodology used in analyzing the past and present cyber-attacks mentioned in the previous section, then in section V, mathematical analysis is carried out to identify the pattern which the cyber-threats and attacks have taken. Finally, section VI discusses the result of the analysis done in the previous section, bringing to light the relationship between the cyber-attacks and predicting the form in which cyber-threats and attacks will take in the future [3].

## II. LITERATURE REVIEW

### 2.1 What is a Threat?

Cyber threat is any act, actor or adversary [4] that can exploit the vulnerabilities in a computer system to cause harm to human life, machines, operations, information and the environment.

### 2.2 What is an Attack?

Cyber-attack is an activity done using a computer to breach the confidentiality, integrity and availability of information and services provided by another computer [5]. Cyber-attacks can be grouped into un-targeted and targeted attacks.

#### 2.2.1 Un-targeted attacks:

Un-targeted attacks are perpetrated randomly without any particular victim in mind [6]. Examples of un-targeted attacks are listed below:

- Phishing
- Scanning
- Water-holing
- Ransomware

#### 2.2.2 Targeted attacks:

Targeted attacks are aimed at a particular victim [7]. Examples of targeted attacks are:

- Industrial espionage
- Denial of Service attack
- Spear phishing/Whale phishing
- Password attack
- Web Application attacks (buffer overflows)

### 2.3 Stages of an Attack

There are four stages involved in the cyber-attack process, namely: Survey, Delivery, Breach, and Affect [4].
Survey: gathering information about a system to discover its vulnerabilities
Delivery: accessing a position where a system's vulnerability can be exploited.
Breach: exploiting vulnerability in a system to gain illegitimate access into it.
Affect: causing harm to a system or with a system which has been attacked.

### 2.4 Threat Agents

These are the adversaries that carry out a cyber-attack. They include the following:

- Individual Crackers
- Insiders: current disaffected employee, former employee
- Business competitors
- Hacktivists
- Terrorists
- Nation States

### 2.5 Impact of Cyber-Attacks

- Financial loss: Illegitimate transfer of funds can be one of the aims of cyber-attack. Other operations in the attack directly or indirectly result in financial loss.
- Lack of Trust: Clients and partners doubt the security facilities available in a business after it has been struck by an attack.
- Loss of Intellectual Property (IP): Intellectual properties like copyright, patents, designs, trademarks and trade secrets are major targets of cyber-attacks. When IP is lost, competitive advantage and revenue suffer loss too [8].

- Loss of sensitive business information: Information loss is a common consequence of cyber-attacks both to individuals and groups. Health information, login credentials, business secrets, and management ideas are among sensitive information lost to cyber breaches [9].
- Business disruption/loss of sales: attacks like denial of service cripple the smooth operation of businesses' services like ordering/purchase and sales/delivery [10].
- Equipment loss: Some malwares destroy hardware they infect, causing the system owners to spend more to replace the destroyed machines [10].
- Devaluation of trade name: The aftermath of Cyber-attacks are characterized with erosion of trade names of businesses as finance and trust are lost [8].
- Loss of customer relationship: Cyber-attacks are responsible for loss of customers of many businesses, resulting in significant loss of profits [11].

## III. SURVEY OF TRENDS

Some popular recorded cyber intrusions perpetrated by individuals and groups from the year 2012 to this present year 2017 are reviewed below:

**a) 2012**

**LinkedIn:** LinkedIn is an employment and business-oriented social network that operates online [12]. In 2012, a hacker named Peace stole about 117 million combinations of emails and passwords from LinkedIn and sold put them for sale on The Real Deal, which is an illegal marketing platform on the dark web.

**b) 2013**

**Houston Astros:** Houston Astros is a baseball team that is based in Houston, Texas, United States of America. When a former staff of St. Louis Cardinals (an American baseball team based in St. Louis, Missouri) called Jeff Luhnow was leaving Cardinals for Houston Astros around December 2011, he handed his work laptop and its password to a staff of Cardinals at that time called Christopher Correa. Correa began to try variations of Luhnow's passwords on Astros's database until he succeeded in getting the password that enabled him access Astros' database called Ground Control [10]. By March and July 2013, Correa scooped player information about how players are obtained and rated.

**Yahoo:** Yahoo is a popular mail service based in the United States. In 2013, data from 1 billion Yahoo accounts were stolen. Such data include user names, phone numbers, and date of birth, password and security questions that are useful for password reset.

**Myspace:** Myspace is an online social network which allows users to submit network of friends, videos, music, blog, profiles, photos, etc. Myspace was intruded in June 2013, by a hacker called "Peace" who stole Usernames and Passwords from about 360 million accounts of users [13].

**Target:** target is a discount store retailer, with headquarters in Minneapolis, Minnesota, United States. Data breach on Target's systems started on the 27th of November, 2013 [14]. The hackers stole 11 gigabytes of data consisting of credit and debit card records of about 110 million customers of Target. The hackers got their way into Target through a refrigeration contractor called Fazio Mechanical. Through phishing email, a variant of the Zeus banking Trojan called Citadel was installed in Fazio's systems. Citadel was used to collect login credentials.

**c) 2014**

**Home Depot:** Home Depot is a retailing company that deals in home improvement supplies like tools, construction materials and services in the United States. Around April to September 2014, Home Depot systems were compromised and 53 million email addresses, together with 56 million credit and debit card details were stolen [15]. Documents from Home Depot show that the company refused to activate intrusion detection feature on its security software-the managers wanted to reduce cost and service downtime, even though that led to insecurity. The feature was specifically made to identify attacks on registers. An investigator said the attack was on the store's registers [16]. Home Depot says a third-party vendor's stolen login credentials was used to gain access to the retailer's network.

**Yahoo:** By late 2014, Yahoo was struck again. This time, 500 million user accounts were compromised and sensitive information stolen includes user names, phone numbers, and date of birth, encrypted password and unencrypted questions that are useful for password reset. Yahoo thinks the hackers were backed by a nation-state [17].

**Anthem medical data breach:** Anthem Inc. is a for-profit health insurance company based in the United States. Hackers breached the personal information of up to 80 million people stored in Anthem's servers. Information stolen includes names, Social security numbers, birthdays, street addresses, employment details and salary of customers and staff of Anthem [18]. The start date of the break is reported to be February 18, 2014. Some security experts suspect China of carrying out the breach, since the country's hackers have frequently hacked health care companies. It is feared that the information stolen could be used for identity theft and making of false insurance claims.

**Premera Blue Cross:** Blue Cross is a non-profit health insurance company based in Washington, United states. Premera was hacked on May 5, 2014, and the data of 11 million people was exposed, [19]. Information accessed by the hackers includes clinical information, residential address, emails, and dates of birth, Social Security Numbers, bank accounts, and insurance claims.

**JPMorgan Chase:** In April 2014, the mutual fund was hit and 83 million customers' information was stolen – data from 76 million households and 7 million businesses [20]. The bank announced that contact information such as names, phone numbers, postal addresses and emails were stolen, thereby increasing the risk of phishing attacks being targeted at the victims of the hack, using their stolen data. Other information such as user IDs, passwords, Social Security Numbers, and account information were said to have remained unexposed to the hackers.

**Sony Pictures:** In November 2014, Sony began to suffer in the hands of a group of hackers called #GOP (Guardians of Peace) who used the internet to leak unreleased movies, staff emails and salaries they had stolen from Sony. #GOP said they were punishing Sony for making a movie (the movie is titled 'The Interview') on the North Korean Supreme leader Kim Jong Un [21]. Sony says the North Korean government backed the attack. US Office of Personnel Management (OPM): The OPM is the US government's department of Human resources charged with employment, and promotion of civil servants and management of benefits and pensions of Federal staff members. On April 25, 2014, hackers registered a fake website (opmsecurity.org) which was used to grab traffic from OPM's digital network [22]. The breach was done using a file called mcutil.dll which contained malware that helped the hackers to access OPM's servers. The malware infiltrated about 15,000 machines.

### d) 2015

**Ashley Madison:** Ashley Madison is a dating website that is meant for people who are married or are in a committed relationship. The site is owned by a Canadian based company called Avid Media. In 2015, a hacking group named 'Impact Team' gave Avid Media 30 days to close Ashley Madison but when the company refused to comply the hackers leaked details of 32 million of the site's customers online [23]. The sensitive information leaked includes transactions, credit card data, emails and user profiles. Impact Team was angry with Ashley Madison for arranging dates for married people and its high charge given to people who want all their data deleted from the site [24].

**Hacking Team:** Hacking Team is an intelligence organisation that develops and sells hacking tools to various governments. In 2015, a hacker called Phineas Phiser. After exploiting vulnerability in an embedded device, the hacker could access an unsecured database in Hacking Team which he also breached. The hacking was a punishment given to Hacking Team for making tools which governments used to hack and spy on people [25].

**Kaspersky Lab:** Kaspersky Lab is a top anti-malware-producing company based in Russia. Kaspersky was struck by hackers whose intent was to steal trade secrets on Kaspersky's latest technology. Kaspersky admitted that the attacker had access to only data that was not critical to its operations [26].

Kaspersky links the attack to the hackers who used the Trojan dubbed Duqu to infiltrate system in India and Belgium in 2011.

### e) 2016

**Federal Reserve Bank of New York:** In February 2016, a group of hackers called Lazarus hackers operating with a malware called Dridex, used the SWIFT code of Bangladesh Bank to instruct the Federal Reserve Bank of New York to transfer $81 million from Bangladesh Bank's account to five different accounts in the Rizal Commercial Banking Corporation (RCBC) in the Philippines. Security experts claim that North Korea is responsible for the attacks [27].

**Democratic National Committee (DNC):** The DNC is the governing body of the US' Democratic Party. On July 22, 2016 wiki leaks published 19,252 emails and 8,034 attachments stolen from the DNC by a hacker who goes by the name "Guccifer 2.0". On November 6, 2016, Wikileaks released more 8,263 emails gotten from the DNC by hackers. The leaked documents indicated that the DNC members were inclined to helping Hillary Clinton win the elections.

Security experts claim the hackers are backed by Russian government [28]. It is said that Russian was interested in doing the Democratic Party down in order for Donald Trump to win the presidential elections.

**Ukraine's Power Grid:** In December 2016, a group of Russian hackers called Fancy bear hacked Ukraine's power grid, putting out 200 megawatts of power, and they also tracked Ukraine's military unit, retrieve communications and location data [29]. Fancy Bear is linked to the breach of the US Democratic National Committee's emails in 2016.

**f) 2017**

World-wide attack: In May 2017, a virus dubbed WannaCry infected about 300,000 computers in about 150 counties. The virus encrypted files in the host computers and asked the owners of the files to pay a ransom in bit coin within a stipulated time (usually seven days) or lose their files [30]. Researchers in Symantec, Kaspersky and South Korea's Hauri Labs claim the older version of WannaCry shares some similarities with the malware used by Lazarus Group (a group alleged to be run by North Korea) to siphon money from Bangladesh Bank's account at the Federal Reserve Bank of New York [9].

## IV. METHODOLOGY

The methodology for comparative survey adopts systematic review which is aimed at providing the trends in cyber-threats and attacks in order to recommend a research direction [31]. The results of the survey would help us to identify and map research areas related to cyber-threat and attack that need to be intensified and possible research gaps. The process for the study is presented in Figure 1, and consists of six process steps and outcomes [32-34].

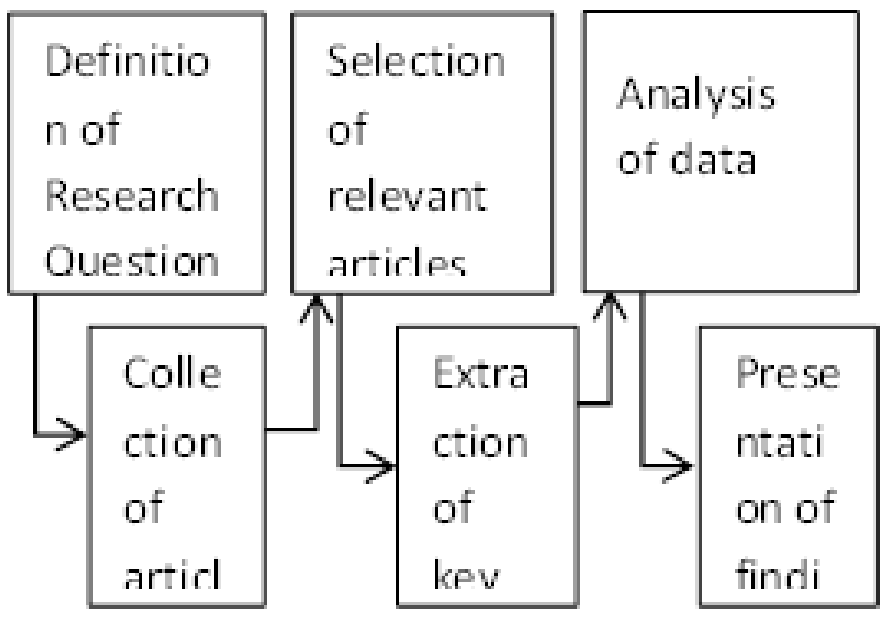

**Figure 1: Research methodology.**

### 4.1 Definition of Research Question

The first stage of the systematic mapping process is the definition of the research questions [35]. The goal of this study was to provide trends in cyber-threats and attacks; we therefore defined four research questions:

- What types of threat and attack was rampant in a particular year
- What system vulnerability caused this type of attack?
- What is the method and outcome of attack?
- Is there any proposed solution? If there is, what is the feasibility of the solution?

### 4.2 Collection of Articles and Paper

In the second stage, we collected relevant articles and papers based on the topic of interest; cyber-threat and attack. The sources of the papers and articles ranges from published journals, online articles, term paper [36].

### 4.3 Selection of Articles

Having collected articles, we selected those ones we care about based on the year published which is 2010 and above.

### 4.4 Extraction of Key Data

The third stage is the extraction of key data from the carefully selected articles. The aim of this stage is to extract relevant data from the article and present the data in a structured form for better understanding [37]. Some of the data items considered are title of article, incident description, date occurred, attack method, application weakness, outcome, cost, attack source geography, items leaked, etc.

### 4.5 Data Analysis

We carried out the analysis of the extracted data in order to show persistent threats and attacks across years. The aim of doing this is to make recommendation of research directions on cyber threats and attacks [38].

### 4.6 Presentation of Result

The last stage is the presentation of findings and recommendation of research area on cyber threats and attacks [39].

## V. COMPARATIVE ANALYSIS

We present the comparative analysis of cyber-threat and attacks using the data collected from www.hackmageddon.com. Three factors are considered in data organisation: motivation, method and target from 2012-2017.

| Motivation | Rate | | | | | |
|---|---|---|---|---|---|---|
| | 2012 | 2013 | 2014 | 2015 | 2016 | 2017 |
| **Cyber Crime** | 57.275 | 49.16667 | 62.3 | 67 | 72.1 | 77.4 |
| **Hacktivism** | 36.3333 | 42.08333 | 24.2 | 20.8 | 14.2 | 4.7 |
| **Cyber Warfare** | 3.45833 | 3.083333 | 11 | 9.8 | 9.2 | 14.5 |
| **Cyber Espionage** | 2.93333 | 5.5 | 2.5 | 2.4 | 4.3 | 3.4 |
| **Art?** | 0 | 0.083333 | 0 | 0 | 0.2 | 0 |
| **N/A** | 0 | 0.083333 | 0 | 0 | 0 | 0 |

**Table 1: Attack motivation comparison table.**

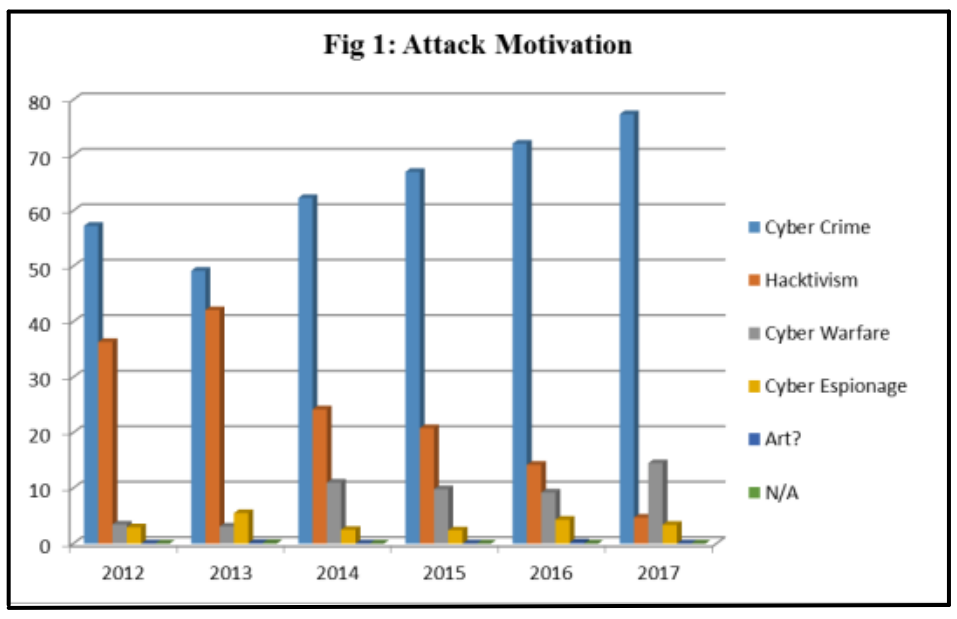

**Figure 2: Attack motivation.**

| Method | Rate | | | | | |
|---|---|---|---|---|---|---|
| | 2012 | 2013 | 2014 | 2015 | 2016 | 2017 |
| **Account Hijacking** | 2.4 | 10.3 | 9.8 | 8.3 | 15.1 | 15.5 |
| **DDOS** | 19.4 | 19.9 | 9.3 | 9.3 | 11.2 | 4.2 |
| **Defacement** | 6.4 | 14.4 | 14.8 | 12.4 | 4.5 | 2.4 |
| **DNS Hijacking** | 1 | 1.7 | 1 | 0.9 | 0.4 | 1.2 |
| **Malicious IFrame/JS** | 0 | 0.8 | 0.7 | 1.1 | 0.4 | 0 |
| **Malvertising** | 0 | 0 | 0 | 2.1 | 1.8 | 1.3 |
| **Malware/POS Malware** | 1 | 2 | 9.4 | 6.4 | 8 | 29.8 |
| **Others** | 12 | 5.6 | 8.9 | 6.7 | 2.9 | 4.2 |
| **SQLi** | 30 | 18.2 | 12.8 | 17.5 | 8.3 | 0.8 |
| **Targeted Attack** | 1.7 | 6 | 9.5 | 10.5 | 11.2 | 15.2 |
| **Unknown** | 23 | 19.9 | 23.3 | 23.9 | 32.8 | 22.3 |
| **Vulnerability** | 2.3 | 0.7 | 0 | 0.8 | 3.4 | 3.1 |
| **XSS** | 0.8 | 0.5 | 0.5 | 0.1 | 0 | 0 |

**Table 2: Attack method comparison table.**

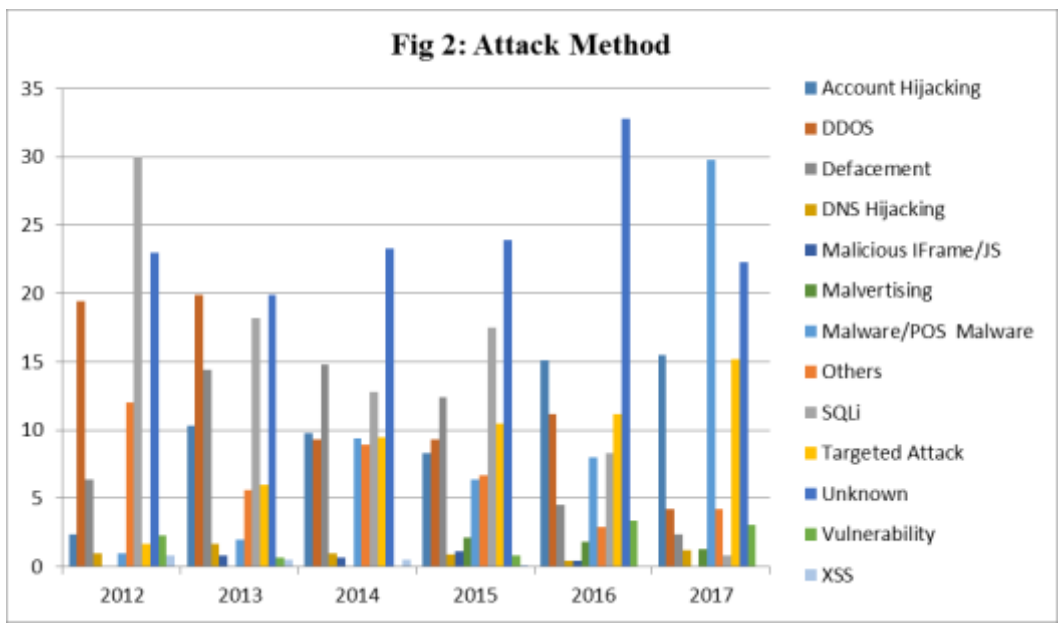

**Figure 3: Attack method.**

| **Target** | **Percentage** | | | | | |
|---|---|---|---|---|---|---|
| | **2012** | **2013** | **2014** | **2015** | **2016** | **2017** |
| >1 | 0 | 1.9 | 3.3 | 4.4 | 4.7 | 7.9 |
| Cryptocurrency Exchange | 0.1 | 0.9 | 0 | 0 | 0 | 2.5 |
| Education | 8.3 | 5 | 6 | 6.9 | 3.4 | 6.8 |
| Finance | 4.3 | 11 | 2.5 | 2.6 | 4.2 | 3.5 |
| Government | 21.3 | 24.3 | 20 | 13.7 | 11.9 | 12.5 |
| Healthcare | 0.8 | 1 | 3.1 | 2.6 | 4.1 | 6.8 |
| Industry | 15.6 | 21.5 | 33.8 | 25.2 | 24.8 | 22.4 |
| Military | 2.4 | 1.6 | 0 | 0 | 0 | 1.4 |
| News | 3 | 5.7 | 2.8 | 2.9 | 1.9 | 0 |
| Online Services | 5.3 | 1.6 | 2 | 2.5 | 3.2 | 0 |
| Organisation | 6.9 | 7.1 | 9.4 | 8.3 | 8.3 | 4.5 |
| Others | 32 | 13.9 | 11.4 | 26 | 24.2 | 9.4 |
| Single Individuals | 0 | 4.5 | 5.7 | 4.9 | 9.3 | 22.3 |

**Table 3: Attack target comparison table.**

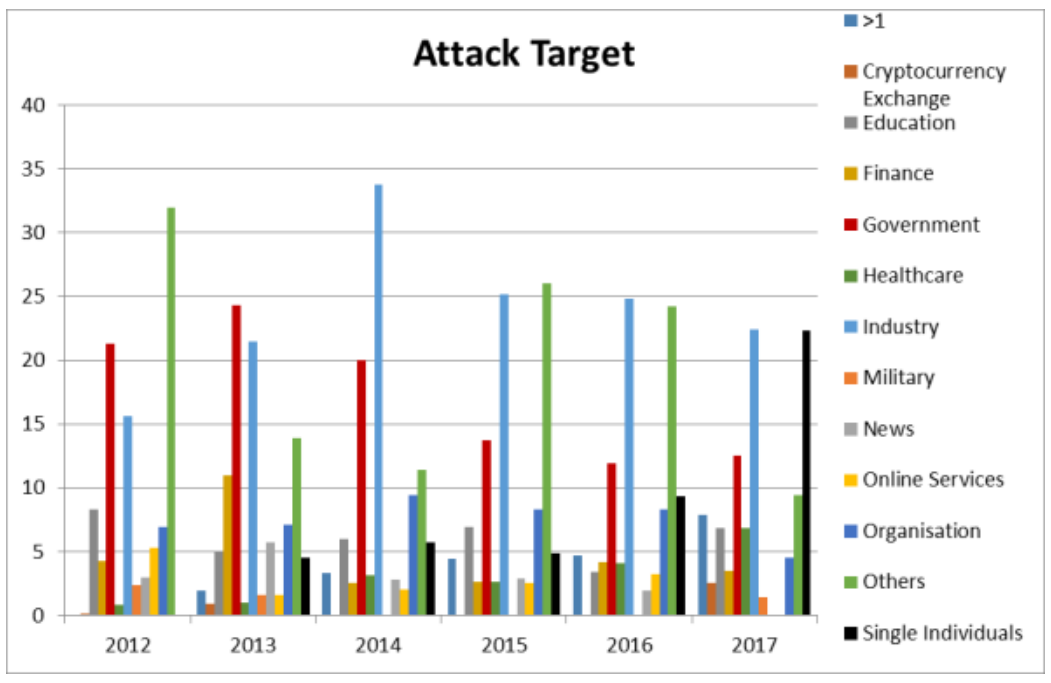

**Figure 4: Attack target.**

## VI. DISCUSSION

### 6.1 Attack Motivation

Our analysis in Figure 2 shows that cyber-crime rate is the highest in 2012 with 57.275 followed by Hacktivism with 36.3333; the same is maintained in 2013 with 49.16667 and 42.08333 for cyber-crime and hacktivism rates respectively Table 1. The cyber-crime rate continues to increase from 2014 to 2017 consecutively with 62.3, 67, 72.1, and 77.4 respectively while hacktivism rate decreases in 2014 to 2017 consecutively with 24.2, 20.8, 14.2, and 4.7 respectively. The analysis indicates increase in cyber warfare rate in 2014 to 2017 with 24.2, 20.8, 14.2 and 4.7 respectively placing it in second position to cyber-crime in 2017. Other motivations like espionage experience no significant change across the years. The point is that cyber warfare simply can mean the use of cyberspace to gain military, political and economic advantage over an opponent but Cyber Crime can be alluded to as a civilian act, highly motivated by criminal benefits like in conventional crime [40]. Cyber-crime is very wide in scope as it covers all malicious acts; hence Figure 2 shows that cyber-crime is the major motivation behind cyber threat and attack since 2012 up to 2017 and the high margin to other motivations indicates its predominant motivation in future. Therefore, more people may likely engage in cyber-crime than other motivation in future. Cyber-warfare on the other hand happens among nations, so the rate may continue to rise as more nations tussle for powers, which indicates its second position to cyber-crime in the future.

### 6.2 Attack Method

Figure 3 shows the method of attacks considered in this survey. The dominant method in 2012 is SQLi with 30 rate followed by unknown attacks (23) and DDOS (19.4) respectively. In 2013, both unknown and DDOS were the most methods applied with equal rate of 19.9 while SQLi method dropped to 18.2 and defacement increased to 14.4 rate. Unknown (23.3) category remained the leading method in 2014, defacement (14.8) rate remained consistent, SQLi rate reduced to 12.8 while Account Hijacking, DDOS, Malware/POS Malware, Others and Targeted Attack maintained approximately equal rate of 9.8, 9.3, 9.4, 8.9 and 9.5 respectively. The major attack methods in 2015 are unknown (23.9), SQLi (17.5) and defacement (12.4) while Targeted Attack (10.5) DDOS (9.3) and Account Hijacking (8.3) attack methods maintained lower rate Table 2. In 2016, Unknown (32.8) method remained the dominant method of attack while Account Hijacking (15.1) rate gained a significant increase. In the same year, DDOS and Targeted Attack are of equal rates of 11.2 while others like SQLi and Malware rates reduced drastically. The dominant attack methods in 2017 are Malware/POS Malware (29.8) and Unknown (22.3) followed by Account Hijacking (15.5) and Targeted Attack (15.2) respectively, others remained very low. Finally, the reason why some attack methods are reducing over the years is the fact that on a regular basis, organizations and individuals keep developing policy, acquiring education and technology as security mechanisms to prevent and mitigate attacks (Michael and Herbert). However, the steady and high rate of Account Hijacking between 2013 and 2015 and its surge in 2016 and 2017 (Figure 3) made it one of the dominant methods and indicate a possibility of future dominance.

### 6.3 Attack Target

Figure 4 shows that the rate of attacking unidentified others are the highest in 2012 with 32 while government and industry follow with 21.3 and 15.6 respectively. In 2013, government was attacked most with the rate of 24.3, followed by industry with the rate of 21.5 while unidentified others reduced to 13.9. The target rate on industry increased to be the highest in 2014 with 33.8 followed by government which reduced to 20 while target rate on unidentified others reduced to 11.4. In 2015 and 2016, both the unknown others (26 and 24.2 rate) and industry (25.2 and 24.8 rate) maintain the highest targets in both years respectively. This is followed by government with 13.7 and 11.9 rates in both years respectively. The Figure 4 shows that the rate of attack on single individual shot up between 2016 to 2017 from 9.3 to 22.3 alongside industry (22.4) attack, making both approximately the predominant targets in 2017. Also, the rate of attack on Finance which has been periodically minimized from 2012 to 2017 may be as a result of security measures put in place by financial institution. Consequently, due to stable rate of attack on industry since 2012 to 2017, we therefore predict its predominance in the future Table 3. Furthermore, attacks on single individuals would continue to surface while that on government would likely increase due to espionage.

## VII. CONCLUSION

Since 2012 up to 2017, cyber-crime has been the major motivation behind cyber threat and attack compared to other motivations. This trend is likely to continue in the nearest future at least. Cyber warfare will follow after cyber-crime as nation's battle each other for supremacy. The constant high rate of Account Hijacking between 2013 and 2015 and its surge in 2016 which was maintained in 2017 indicates a possibility of future dominance. Consequently, the stable rate of attack on industry since 2012 to 2017 indicates its likely predominance in the future. Also, attack on single individuals and governments would likely increase due to espionage.